# Agile Governance Theory: conceptual development

*Alexandre J. H. de O. Luna (Federal University of Pernambuco – UFPE, Pernambuco, Brazil) - ajhol@cin.ufpe.br*

*Philippe Kruchten (The University of British Columbia – UBC, British Columbia, Canada) - pbk@ece.ubc.ca*

*Hermano P. de Moura (Federal University of Pernambuco – UFPE, Pernambuco, Brazil) - hermano@cin.ufpe.br*

**ABSTRACT**. **Context**: Competitiveness is the key to a sustainable development and it demands agility at the business and organizational levels, which in turn requires a flexible and customizable IT environment and effective and responsive governance in order to deliver value to the business. **Objective**: This paper describes the conceptual development of a theory for analyze and describe agile governance in order to increasing the success rate of their practice, achieving organizational performance and business competitiveness. **Method**: We adopt a multi-method research, framing the theory conceptual development using Dubin's method of theory building. **Results**: We have developed a conceptual framework of the theory encompassing its constructs, laws of interaction, boundaries and system states. **Conclusion**: This theory can provide a better understanding of the nature of agile governance, by mapping of its constructs, mediators, moderators and disturbing factors, in order to help organizations reach better results.

**Research Support.** *The authors acknowledge to CNPq, CAPES, ATI-PE, PGE-PE, Brazil's Science without Borders Program, UBC and UFPE by the research support.*

**TÍTULO: Teoria da Governança Ágil: Desenvolvimento Conceitual.**

**RESUMO**. **Contexto**: A competitividade é a chave para um desenvolvimento sustentável e exige agilidade tanto no nível do negócio quanto em nível organizacional, que por sua vez requer um ambiente de TI flexível e personalizável, bem como uma governança efetiva e responsiva, a fim de agregar valor a este negócio. **Objetivo**: Este artigo relata o desenvolvimento conceitual de uma teoria para analisar e descrever governança ágil, a fim de aumentar a taxa de sucesso de sua prática, alcançando desempenho organizacional e competitividade nas organizações. **Método**: Este trabalho adotou uma abordagem de pesquisa multi-método, enquadrando o desenvolvimento conceitual teoria por meio do método de construção de teoria de Dubin. **Resultados**: Foi desenvolvido um *framework conceitual* da teoria, englobando: seus construtos, as leis de interação que regem suas relações, suas fronteiras e os estados do sistema derivados destes componentes. **Conclusão**: Esta teoria pode proporcionar uma melhor compreensão da natureza da governança ágil, através de mapeamento de seus construtos, mediadores, moderadores e fatores de perturbação, a fim de ajudar as organizações a alcançar melhores resultados.

**Palavras-chave.** *Sistemas de Informação, Governança Ágil, Gerenciamento de TI, Gerenciamento de Projetos, Engenharia de Software.*

## 1   Introduction

As stated in The Global Competitiveness Report 2011-2012 elaborated by the World Economic Forum (2011), the world economy moved in 2011 around US$80.33 trillion in GDP (PPP[1]). In keeping with IMF (2012), at exchange rates, the economic output of the world is expected to expand by US$28.7 trillion from 2010 to 2017. In addition, the New York Stock Exchange (NYSE) is a stock exchange where the largest companies in the world, which are responsible for producing most of the wealth generated by those mentioned countries, negotiate their capital. The market capitalization of the NYSE listed companies, encompassed US$14.24 trillion as of December 2011, as well had as average daily trading value approximately US$153 billion in 2008 (WFE, 2013).

Undoubtedly, this is a very competitive context where the decisions should be made sometimes without the complete information required, as well as they should be communicated to the relevant sectors of the organization, which must have the capability to respond and redirect their actions to these changes in a wide and coordinated manner. Any mistake might costs millions of dollars or even can cost the business survival. Indeed, improving the competitiveness of governments and companies should result in significant economic outcomes.

Competitiveness seems related to make more, better and faster, with less resources (Janssen & Estevez, 2013). At the same time, governance is closely related with the ability to steer (to guide, to govern) an organization, which may be a company, a government or a society (Bloom, 1991). In other words, governance is a key driver to "make things happen" on organizational environment. Also, "to be" and "to look" is deeply related with transparency in decisions, actions and results of an organization, something closely related with governance. These thoughts would guide us to imply that the way to competitiveness pass by the application of a "good governance" (UNESCAP, 2013; World Bank, 2006).

In this context, the information and communication technologies (ICT or IT) are the link between the decision-making ability, the willingness strategic, and the competence to put into practice these tactics concretely. In this scenario, *IT governance*, through which *corporate governance*[2] is applied, has emerged as an option to the effective management and control of IT services in organizations (IT Governance Institute, 2001).

In addition, the design and maintenance of the IT systems for enterprise agility are challenging when the products and services must be compliant with several regulatory aspects (often needing to be audited) (Wright, 2014). The establishment of the necessary management instruments and governance mechanism to fulfill this mission passes by the application of models and frameworks that many times have no guidance details of how to implement and deploy them (such as ITIL and COBIT, among others), affecting the organizational competitiveness (Gerke & Ridley, 2009; Mendel, 2004).

---

[1] Purchasing power parity (IMF, 2012).
[2] "It is the set of processes, policies, rules, laws and institutions that affecting the way as a corporation is directed, administered or controlled" (Cadbury, 1992).

Consequently, the challenges become even greater when dealing with these matters in a global software development and distributed environment, where cultural differences, awareness and communication style, if not treated properly, can lead to conflicts. Arguably, in Global Development Environments deal with governance is even more relevant, as well as implementation of governance mechanisms an issue even greater challenging (Dubinsky, Ravid, Rafaeli, & Bar-Nahor, 2011).

In fact, governance is a cluster of steering *capabilities*[3], based on three dimensions: (1) plan strategically; (2) establish mechanisms to ensure accomplishment of the strategic planning; and, (3) sense and respond to change. In turn, every dimension has its respective concepts, actions and analogies, as depicted in **Fig. 1**.

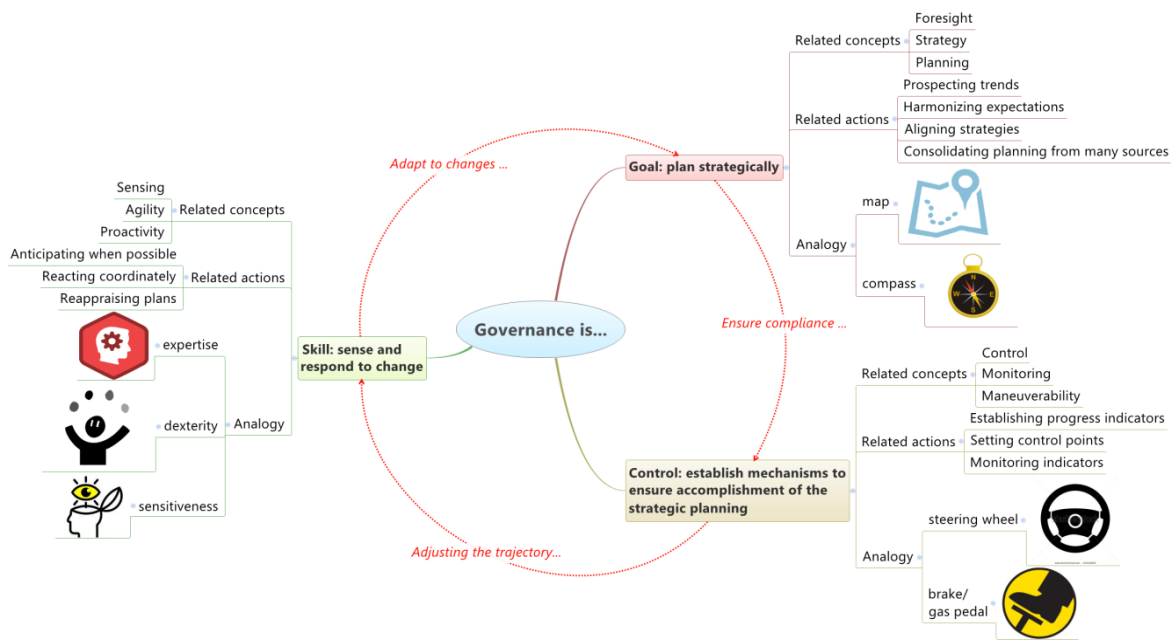

**Fig. 1.** Governance dimensions and analogies. **Source**: Own elaboration.

As stated by Luna, Kruchten, Pedrosa, Almeida Neto, & Moura (2014), chronologically, *agile governance* has been proposed by some authors (Cheng, Jansen, & Remmers, 2009; Luna et al., 2010; Luna, Kruchten, & de Moura, 2013; Luna et al., 2014; Qumer, 2007). At the same time, its concept has been evolved over time, in which its first two definitions (Cheng et al., 2009; Qumer, 2007) were focused in *agile software development*, whereas its third definition Luna et al. (2010) has proposed a *wide application of principles and values of agile software development* (Beck et al., 2001) *to the conventional governance processes*. Recently, Luna et al. (2013, 2014) have proposed a concept of agile governance for encompass the wide and multidisciplinary nature of the phenomena related. In addition, Luna (2009), has developed a framework for agile governance, in order to implement and improve governance in organizations, called MAnGve. This framework is focused to the deployment process, as a catalyzer to accelerate the deployment of governance. The MAnGve framework is designed to mitigate the lack of practical focus found in conventional governance models (MAnGve.org, 2009). However, altogether the agile governance phenomena still remained unexplored in depth.

---

[3] The term *"capability"* regards to a feature, *faculty* or process that can be developed or improved (Vincent, 2008).

Based on those motivation, arises as a relevant issue the understanding of the agile governance phenomena and the contexts in which they occur. Once the agile governance phenomena are better understood in their essence, map their constructs, mediators, moderators and disturbing factors from those phenomena in order to help organizations to achieve better results in their application: reducing cost and time, increasing the quality and success rate of their practice. This work has a focus on organizations that need to operate (sense and respond) in turbulent and/or competitive environments, as well as that need to grow sustainably, reacting as a coordinated whole, attaining greater enterprise agility and supporting their overall strategy, in the context of IT Governance.

In the following sections we will describe the methodological approach adopted to conduct this research (Section 2), the conceptual development of the theory, and its results, in Section 3. At Section 4, we will conclude and present implications for research and practice.

## 2 Methodology

As reported by Creswell (2003), a researcher should make use of a framework to guide his or her project research since the identification of the epistemological stance that underpins the researcher's philosophical stance, until the procedures for collecting and analyzing data. According to Myers (1997), the relevant items that should be considered in the research project are: (1) philosophical perspective, (2) methods, (3) techniques of data collection, and (4) methods of analysis and interpretation of data; similarly to those proposed by Creswell (2003). Using as references the views of Myers (1997) and Creswell (2003), and inspired by some study designs applied by researchers who we had contact over time, we have elaborated a research framework depicted in **Fig. 2**, which treats the relevant aspects to be considered by this study.

This type of research can be classified as multi-method or mixed (Creswell, 2003) where we apply in combination a systematic literature review, social network meta-ethnography and semi-structured interviews with an emphasis on qualitative aspects; and the cross-sectional research explanatory survey with quantitative approach. Our position is that theories should be useful, and, whenever possible, practical and applicable in essence! In keeping with Sjøberg, Dybå, Anda, & Hannay (2008) we adopt the view of the philosophical school of pragmatism, considering both specific beliefs and methods of inquiry in general should be judged primarily by their consequences, by their usefulness in achieving human goals. According to this philosophical perspective, the meaning of an idea corresponds to the set of its practical implications (James, 1995).

We have assessed the following theory-building methods: (1) Dubin's Theory-Building Method (Dubin, 1978); (2) Grounded Theory-Building (Corbin & Strauss, 1990); (3) Software Engineering Theory-Building Framework (Sjøberg et al., 2008); and (4) Lynham's General Method (Lynham, 2002b) — against the selection criteria: strengths, limitations, and completeness. This analysis revealed that Dubin's Theory Building Method was best suited for this study in combination with some techniques from Grounded Theory.

Our research had two major phases: (1) the theory emergence; (2) the theory assessment. This paper is focused in the description of the **Phase 1** of this research, specifically in the **stage 2** of the **Fig. 2**.

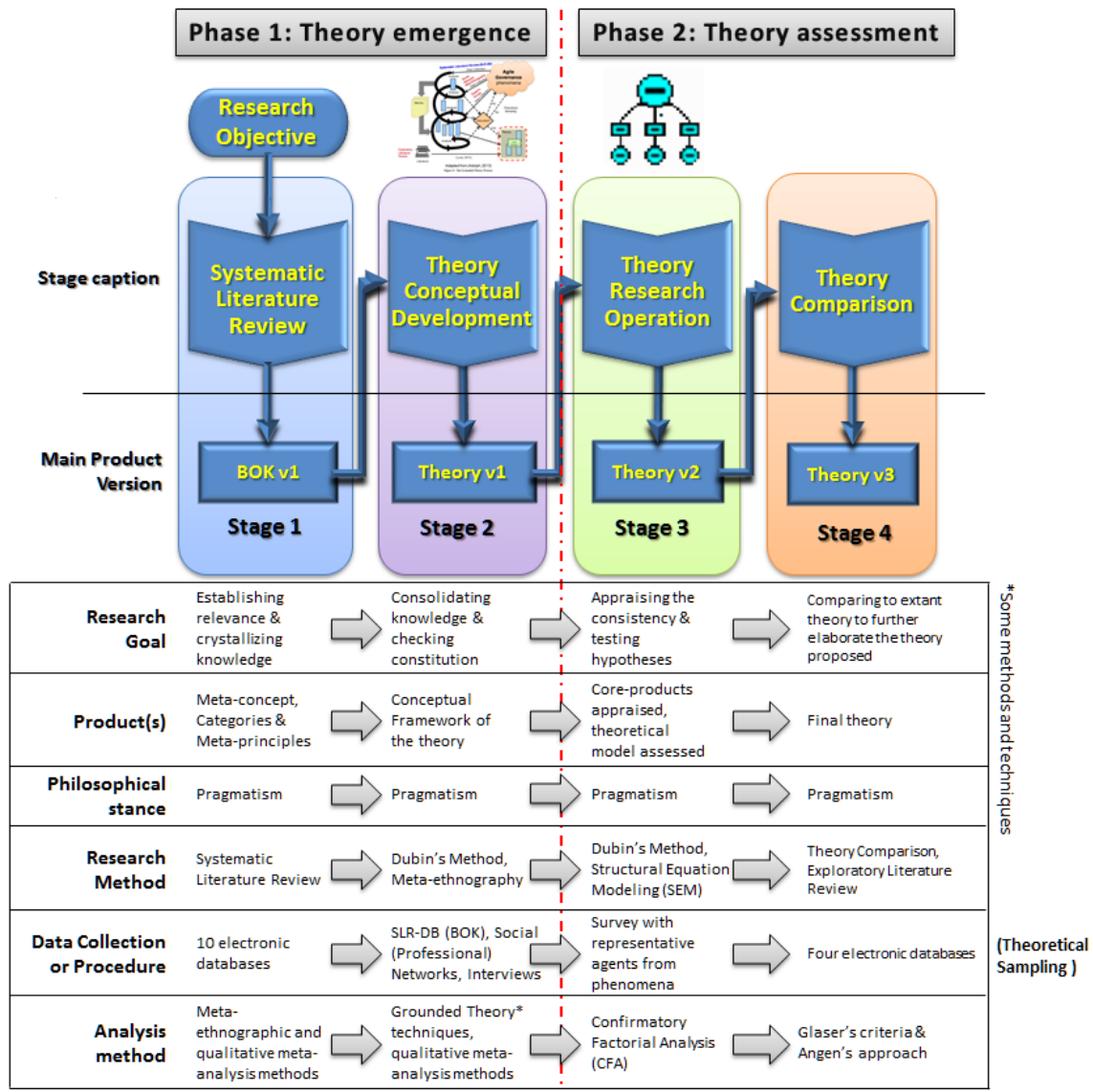

**Fig. 2.** Research framework. **Source**: Own elaboration, inspired from (Adolph, Kruchten, & Hall, 2012; Dorairaj, Noble, & Allan, 2013; Monasor, Vizcaíno, Piattini, Noll, & Beecham, 2013).

At this stage we carried out conceptual development of the theory, following the initial four steps that comprise part one of Dubin's methodology for theory building research (Dubin, 1978). At that time, we have identified and characterized the core-components of the emerging conceptual theoretical framework: units (constructs), laws of interaction, boundaries and system states. At stage 1, to complement data from the findings of the systematic literature review published in (Luna et al., 2014), we add two new theoretical sampling sources: (1) an ensemble of social networks composed by researchers and practitioners in governance, management and agile methods (Murthy, 2008; Wolfe, 1997), including 12 professional and research groups related to governance; and, (2) semi-structured interviews with ten representative agents from the phenomena in study, including researchers and practitioners in governance, management and agile methods. In order to analyze and synthetize findings from those sampling, e.g., emerging relations between the categories already identified in the previous stage, and the new categories and connections that can emerge during this stage, we adopted some techniques from Grounded Theory described by (Corbin & Strauss, 1990; Eisenhardt, 1989; Pandit, 1996) and the meta-ethnographic and qualitative meta-analysis methods described by (Britten et al., 2002; Noblit & Hare, 1988).

The first four of Dubin's eight research steps comprise the first part of the theory building research process, which entails conceptual development of the theory (or theoretical model). The steps in this part of the theory-building process include: (1) Identification and definition of the units of the theory (i.e. the elements that interact to create the phenomenon, or constructs); (2) Determination of the laws of interaction that state the relationships between the units of the theory; (3) Definition of the boundaries of theory to help focus attention on forces that might impact the interplay of the units; (4) Definition of the theory's system states (i.e. different situations which may affect the interaction of the theory's units).

The best known graphical representation of the Dubin's method was popularized by Lynham (2002a, p. 243) in her book chapter "*Quantitative Research and Theory Building: Dubin's Method*" in "*Advances in Developing Human Resources*". In fact, the *Fig. 1* from her book chapter conveys the idea that Dubin's method is "linear, sequential" and without refinement cycle. However, after reading the Dubin's book "*Theory building: a Practical Guide to the Construction and Testing of Theoretical Models*" our opinion is that the representation of the Dubin's method proposed by Lynham (2002a) "does not do justice" to the rich description, generously provided by Dubin (1978) in his book.

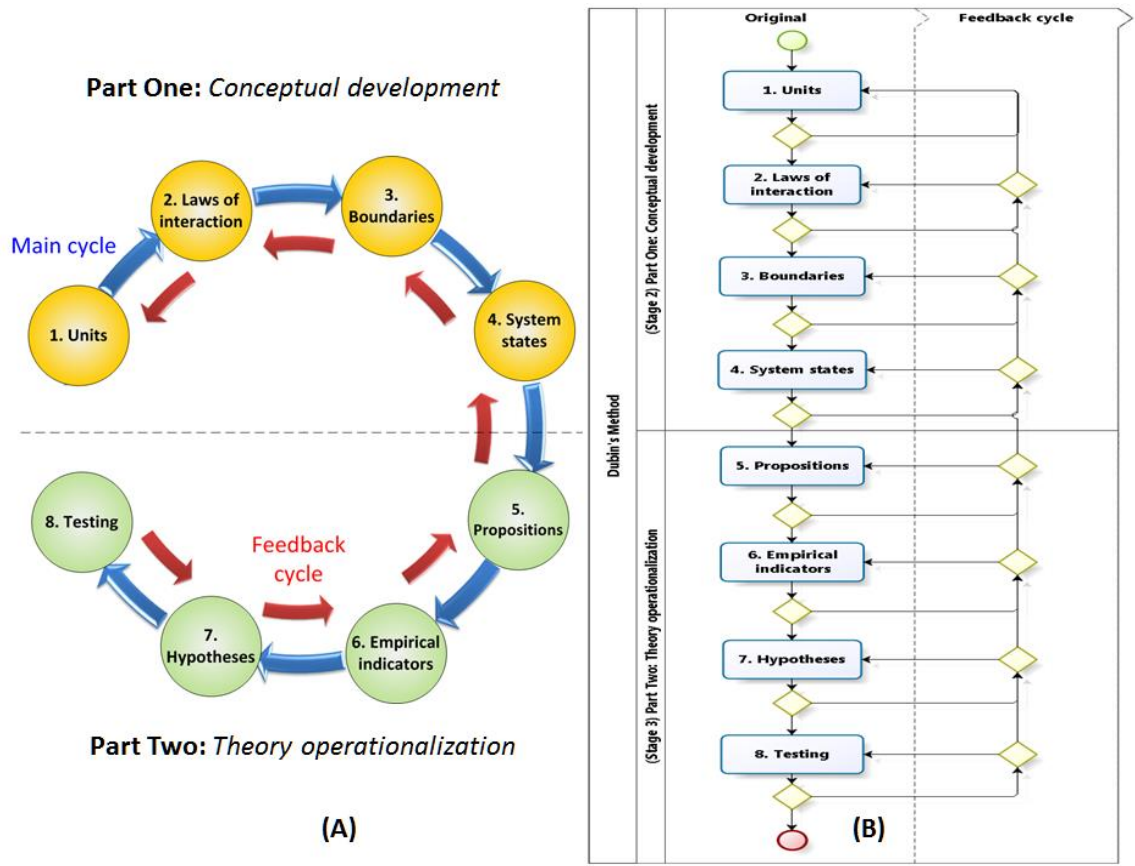

**Fig. 3.** Dubin's method: including feedback cycle. **Source**: Adapted from (Dubin, 1978).

Unfortunately, maybe Dubin has some guilt on that issue, because in none of the 304 pages of his book there is no graphical representation of the method, despite of the eloquent description and abundant number of examples and analogies. So, avoiding discussing that Lynham (2002a) was unhappy in her graphical representation of the Dubin's method, we would like to introduce our own view about the Dubin's method in **Fig. 3**, making explicit the feedback cycle for each step of the method.

## 3   Theory: conceptual development

Considering that: (1) Nowadays agile governance are a poorly explained phenomena (Luna et al., 2014); (2) Currently people apply agile governance serendipitously or facing many challenges (Barton, 2013; Dubinsky & Hazzan, 2012; Luna et al., 2014; Parcell & Holden, 2013); (3) According to Gregor (2006), Bordage (2009) and Edmondson & McManus (2007), a theory or a conceptual framework is an instrument compatible with the stage of development of the phenomena in study, and a significant contribution, which can give a better understanding about them; (4) Improving competitiveness of governments and companies through the improvement of their governance and management shall result in significant economic returns (Porter, 1985; WFE, 2013; World Economic Forum, 2011).

It is assumed that a theory for analysis and description (Gregor, 2006), should be a legitimate classification for the emerging theory from this work, which can be used to describe what agile governance is, as well as help to interpret and understand how agile capabilities and governance capabilities can be applied in order to achieve business agility.

In the following sections we will describe the premises, and the key elements of the emerging theory resulting from the four first steps of the Dubin's research method.

### 3.1   Foundational Premises of the Theory

The agile governance phenomena emerges in the context of organizational environment, as a young and nascent area, eight years old, driving people to apply agile capabilities upon governance capabilities to provide business agility (Luna et al., 2014). The predominant concern of them is to deliver value faster, better and cheaper to the business in sustainable cycles. On the organizational context, governance is the keystone to create the necessary engagement of all units of the organization, attaining greater enterprise agility and supporting its overall strategy.

**Premise 1:** Thus, our proposal introduces agile governance as the application of agility upon the system responsible for sense, respond and coordinate the entire organizational body: the governance (or steering) system. Differently from specific agile approach widely held on organizations (such as agile software development or agile manufacturing), in which the influence is limited to a localized result, usually few stages of the chain value (Porter, 1985) of the organization (Luna et al., 2014).

**Premise 2:** Concerning to positioning of the phenomena, we can imply the agile governance as socio-technical phenomena positioned in a chaordic range between the innovation and emergent practices from agile (and lean) philosophy and the status quo of the best practices employed and demanded by the governance issues. The socio-technical nature of agile governance is substantiated due we are handling with the understanding of the intersections between technical and social aspects: considering people as agents of change in organizations, in contexts where technology is a key element (Luna et al., 2014).

**Premise 3:** Finally, the third premise is the definition of agile governance as a broad concept and its meta-principles, and meta-values proposed in (Luna et al., 2014).

- **Broad concept:** "*Agile governance is the ability[4] of human societies[5] to sense, adapt and respond rapidly and sustainably to changes in its environment, by means of the coordinated combination of agile and lean capabilities with governance capabilities, in order to deliver value[6] faster, better, and cheaper to their core business[7].*"

- **Meta-principles:** We have proposed the following six meta-principles for agile governance, in order to guide future researches and, especially, to drive practices (Luna et al., 2014).

    (i) **Good enough governance:** "*The level of governance must always be adapted according to the organizational context*".
    (ii) **Business-driven:** "*The business must be the reason for every decision and action*".
    (iii) **Human focused:** "*People must feel valued and incentivized to participate creatively*".
    (iv) **Based on quick wins:** "*The quick wins have to be celebrated and used to get more impulse and results*".
    (v) **Systematic and Adaptive approach:** "*Teams must develop the intrinsic ability to systematically handle change*".
    (vi) **Simple design and continuous refinement:** "*Teams must deliver fast, and must be always improving.*"

- **Meta-values:** In order to achieve better results, we believe that teams should use those meta-principles, having the support of meta-values to guide actions, which in turn also can help them to differentiate the approaches of both: conventional and agile governance. As a consequence of this research we have come to value the meta-values from the column "A" of the **Table 1**. That is, while we recognize the value in the items on the right (column B), we value the items on the left more (column A).

**Table 1.** Agile governance meta-values. **Source**: Own elaboration, inspired from (Beck et al., 2001).

| ID | (A) Agile Governance | (B) Conventional Governance |
|---|---|---|
| 1 | It is more about **behavior** and **practice**... than... | ...**process** and **procedures**. |
| 2 | It is more about achieve **sustainability** and **competitiveness**... than... | ...be **audited** and be **compliant**. |
| 3 | It is more about **transparency** and **people's engagement** to the business... than... | ...**monitoring** and **controlling**. |
| 4 | It is much more about **sense**, **adapt** and **respond**... than... | ...follow a **plan.** |

---

[4] A natural or *acquired* skill or talent.

[5] We have tried to encompass any kind of organizations, such as: companies in any industry, non-profit institutions, as well as governments in any level or conjunction (cities, provinces, countries, or even governments associations, e.g. The United Nations).

[6] An informal term that includes all forms of value that determine the health and well-being of the firm in the long run.

[7] Is the *raison d'être* of any organization, the cause of its existence.

## 3.2 Basic Constructs of the Theory

Theory units (or constructs) are the concepts of the theory, or the basic ideas that make up the theory, or "knowledge plots" from which the theory is built, i.e., the building blocks of the theory or the elements that come together in the theory (Dubin, 1978). The units represent the things (or things properties) which the researcher is trying to make sense of and which are informed by literature and experience.

In order to determine the concepts that would be included in the theory, we developed a set of theoretical samplings described in Section 2. At this step, we have identified the units of the theory, whereas during the process of identification of the attributes for each theory unit, they have emerged based on the following criteria of development: i) the application of the constant comparative method of qualitative analysis onto data with emerging categories (Glaser, 1965); ii) the balance between the frequencies of citation of them in the sources of the theoretical sampling chosen; iii) the representativeness desired by the theory design, trying to answer: how well the attributes can describe the construct; and, iv) the ability to translate the key characteristics of relevant meaning about the theory unit; and, finally, v) due the fact that it can be applied in most instances of this theory unit, some of them found by complementary (exploratory) literature review about this topic.

As a result, we have identified six theoretical units (constructs) that can describe and explain agile governance phenomena, by means of their relations, and interactions, namely:

(1) **Effects of environmental factors [E]:** conceptualizes the effects sensed by the organizational context, as a result of the influence caused by the external environment in which the organizational context resides.
(2) **Effects of moderator factors [M]:** conceptualizes the effects sensed by the organizational context as a result of the influence caused by moderator factors forming part of this context. Those factors tend to oppose the organizational performance, i.e., inhibiting or restraining the organizational performance, in turn, retarding its advance. The nature of these factors varies according to the particularity of each organizational context.
(3) **Agile capabilities [A]:** is the ability to acquire, develop, apply and evolve competencies[8] related to principles, values and practices, from agile and lean philosophy on organizational context.
(4) **Governance capabilities [G]:** is the ability to acquire, develop, apply and evolve competencies related to the way as an organizational context is conducted, administered or controlled, including the relationships between the distinct parties involved and the aims for which a society is governed.
(5) **Business operations [B]:** conceptualizes the set of organized activities involved in the day to day functions of the business, conducted for the purpose of generating value delivery.
(6) **Value delivery [R]:** conceptualizes the ability to generate results (and become persistent the benefits arising from them) to the business by means of the delivery of value, whereas includes all forms of value that determine the health and well-being of the organization in the long run.

According to Xu, Zhu, & Liao (2011), **organizational context** is an important factor that significantly affects Information Systems (IS) research and practice, and its effectiveness, as different components of the organizational context constitute different environments in which IS are developed and implemented. Those constructs can be instantiated for the following organizational contexts: (1) teams, (2) projects, (3) business units, (4) enterprise, or even in a (4) multi organizational setting. In this conceptual development, "team" is a generic word that can be applied for several complementary connotations in organizational context, such as: technical people, business people, and even the steering committee. See more details about in Section 3.6.

---
[8] The term "competency" refers to a combination of skills, attributes and behaviors that are directly related to successful performance on the job (Landström, Mattsson, & Rudebeck, 2009).

Dubin (1978) emphasized the importance of characterizing and classifying the nature of units used in a theory. Units, he argued, must be differentiated "*in order to draw out their consequences*" (p. 37). Units can be differentiated by both their properties, which represent dichotomous characteristics (i.e., attribute versus variable, real versus nominal, primitive versus sophisticated, and collective versus member), as well as by their class (i.e., enumerative, associative, relational, statistical, and summative). In short, the application of Dubin's logic on those units clarifies that the units are *variable*, *real*, *sophisticated*, *collective*, and, about the class: *associative*, because they can have a zero or negative values.

### 3.3 Laws of interaction

The laws of interaction describe the interactions that govern the theory, i.e., the synergy between the units of the theory. The laws of interaction presented in this section are statements of relationship that explain how the theory's units are connected, i.e., specify the relationships, or linkages, between the units. According to Dubin (1978), it is these relationships between units with which science is centrally concerned; the scientist's objective is to account for the variance in one unit by specifying a systematic linkage of the unit with at least one other. Dubin labeled the systematic linkages among units within a theoretical model "laws of interaction." He specifically chose the term laws of interaction to "*focus attention on the relationship being analyzed*," (Dubin, 1978, p. 90).

We have identified six laws of interaction for the theory, which statements are depicted as follows.

- **1st Law (of agile governance):** "*Agile governance arises when agile capabilities [A] are combined and coordinated with governance capabilities [G], activating or intensifying an increase in the level of business operations [B], which in turn increases the value delivery [R]*".
- **2nd Law (of specific agile approach):** "*An specific agile approach arises when agile capabilities [A] are applied in different aspects of the organizational context, which are not governance capabilities [G], activating or intensifying an increasing in business operations [B], which in turn increases the value delivery [R].*"
- **3rd Law (of moderator factors effects):** "*There are internal moderator factors whose effects [M] can inhibit or restraining the agile capabilities [A] and governance capabilities [G], or even reduce business operations [B], which in turn decreases the value delivery [R].*"
- **4th Law (of environmental factors effects):** "*There are environmental factors whose effects [E] can disturb the organizational context, influencing: the effects of moderator factors [M], agile capabilities [A], governance capabilities [G] and business operations [B], which in turn affects in some level the value delivery [R].*"

**Please cite this article as:**
Luna, A. J. H. de O., Kruchten, P., & Moura, H. P. de. (2015). Agile Governance Theory: conceptual development. In D. M. G. Sakata (Ed.), *12th International Conference on Management of Technology and Information Systems*. São Paulo: FEA-USP.

- **5th Law (of sustainability and competitiveness):** "*The combined and coordinated coupling of agile capabilities [A] and governance capabilities [G] reduces the effects of environmental factors [E] and the effects of moderator factors [M] upon the organizational context, contributing to decreases the inhibition, restriction or disturbing on organizational context, and decreasing their harmful effects upon business operations [B] over time, which in turn increases the value delivery [R].*"
- **6th Law (of value delivery):** "*Influence on business operations [B] will generate directly proportional effects on value delivery [R].*"

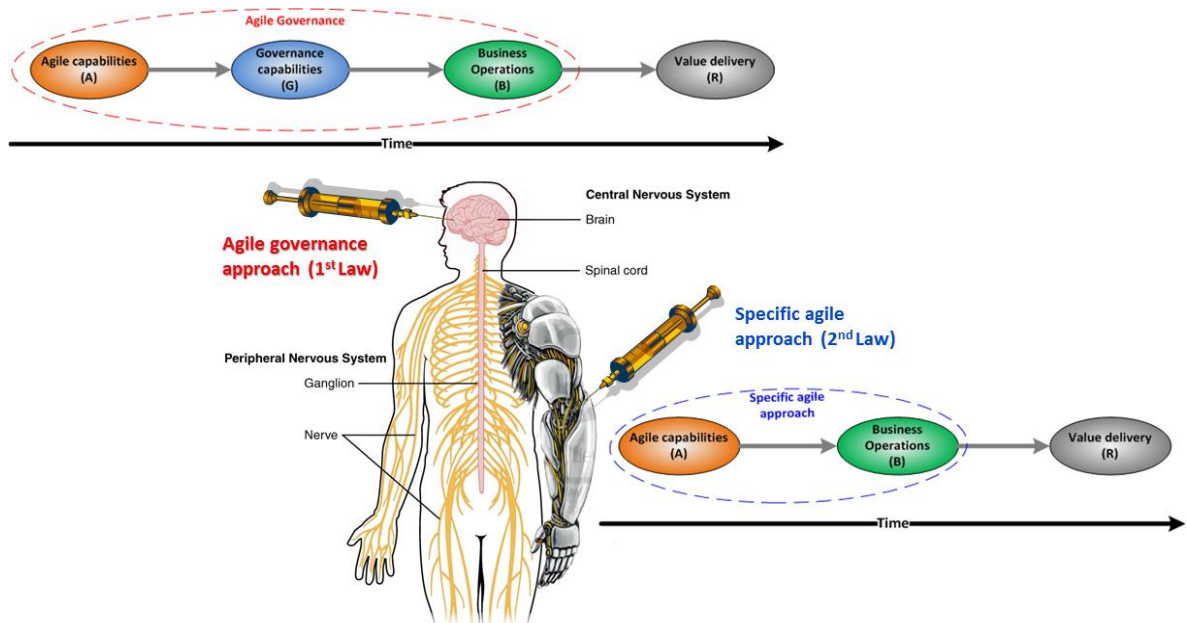

**Fig. 4.** 1st and 2nd Laws of interaction illustrated. **Source**: Adapted from (Luna et al., 2014).

Dubin (1978) highlighted three general categories or types of laws of interaction, namely, *categoric*, *sequential*, and *determinant*. In addition to specifying the three categories of laws of interaction, Dubin (1978) indicated that a law of interaction may have four different levels of efficiency, each of which provides a different level of predictive power and understanding, namely: *presence-absence*, *directionality*, *covariation*, and *rate of change*.

Indeed, every law of the theory is a *sequential* law of interaction at the second level of efficiency (*directionality*), because they are apparent from the inclusion of a time dimension, as well as they describe the directionality of a relationship between two or more units. In order to illustrate the first and second laws of interaction, we produce the **Fig. 4**.

### 3.4 Boundaries

Theories are intended to model some element of the real world. The boundaries of a theory identify which aspects of the real world the theory is attempting to model and which it is not (Lynham, 2002a). Thus, the boundaries of a theory delineate the domains or territory over which the theory is expected to hold true (Dubin, 1978). Both units and laws must comply to the theory's boundary-determining criteria before the theory is complete (Dubin, 1978).

**Please cite this article as:**
Luna, A. J. H. de O., Kruchten, P., & Moura, H. P. de. (2015). Agile Governance Theory: conceptual development. In D. M. G. Sakata (Ed.), *12th International Conference on Management of Technology and Information Systems*. São Paulo: FEA-USP.

It is important to first clarify some basic related concepts, namely, boundary criteria, as follows: (1) **interior boundary-determining criteria**, i.e., are those that are "*derived from the characteristics of the units and the laws employed in the theory*" (Dubin, 1978, p. 128); and, (2) **external boundary-determining criteria**, i.e., are those *"imposed from outside the theory"* (p. 132). The number of boundary-determining criteria also has an influence on the *homogeneity* of the theory's domain. As the number of boundary-determining criteria increases, the theory's units and laws of interaction become more homogeneous. According to Dubin, in fact over the open boundary there is exchange between the domains through which the boundary extends, whereas over the closed boundary, exchange does not take place (Dubin, 1978, p. 126). **Table 2** depicts the summary of classification for the boundaries of the theory.

Table 2. Theory boundaries. **Source**: Own elaboration.

| ID | Boundary type | Boundary | Dubin's homogeneity criteria | |
|---|---|---|---|---|
| B1 | The open boundary | **Organizational contexts**: only those units and laws of interaction that relate to the organizational contexts of the Agile Governance in *IT teamwork* perspective are within the domain of this theory, insofar it is: team, project, business unit, enterprise, or a multi organizational setting. | • Teamwork<br>• Information Technology (IT) | 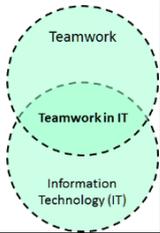 |
| B2 | The closed boundary | **IT Governance domain**: only those organizational approaches that can be classified as *IT Governance*, fall within the domain of this theory. | • Governance<br>• Information Technology (IT) | 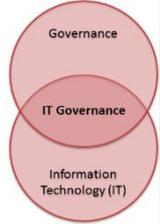 |

After complete this third step of the Dubin method we are able to represent graphically the conceptual framework of theory, as depicted in **Fig. 5**. The constructs Environmental factors' effects ($E_{i_{(1 \to n)}}$)[9] and Value delivery ($R_{j_{(1 \to m)}}$)[10] are border phenomena and they are represented by means of red and black arrows, respectively. The gray arrows connecting constructs describe the interaction between each one of them, stated by the laws of interaction (see Section 3.3).

## 3.5 System states

The system states of the theory represent conditions of the theoretical model in which the units of the theory interact differently. In order to identify the system states of a theory, this theory must first be considered as a system (Lynham & Chermack, 2006). This means that the theory must be perceived as a bounded set of units, interrelated by laws of interactions, from which deductions are possible about the behavior of the overall system (Lynham &

---

[9] The notation describes the fact that each factor from the external environment receives an index "i", which varies from 1 to "n", where "n" is the total number of "environmental factors' effects" [E] that operates in a particular instance of the theory.

[10] The notation describes the fact that each outcome from the organizational context has its "value delivery" [R] component, and receives an index "j", which varies from 1 to "m", where "m" is the total number of outcomes from organizational context, in a particular instance of the theory.

Please cite this article as:
Luna, A. J. H. de O., Kruchten, P., & Moura, H. P. de. (2015). Agile Governance Theory: conceptual development. In D. M. G. Sakata (Ed.), *12th International Conference on Management of Technology and Information Systems*. São Paulo: FEA-USP.

Chermack, 2006). Systems may exist in different states. A system state is a condition of the theoretical model in which the units of the system interact particularly. During these different system states, each of the system units takes on a characteristic value for some time interval (Dubin, 1978). Dubin (1978) further identified three criteria of importance to the researcher-theorist when identifying the system states of the theory, namely, (i) *inclusiveness*[11], (ii) *persistence*[12], and (iii) *distinctiveness*[13].

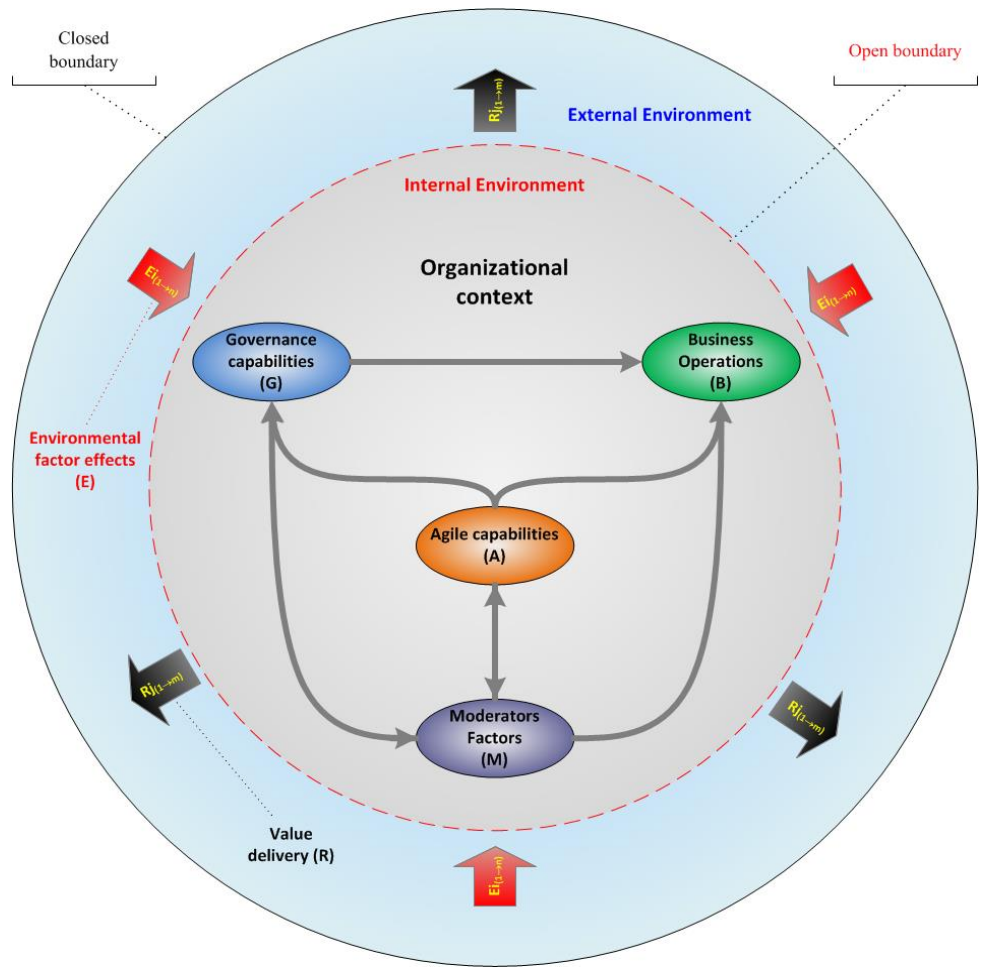

**Fig. 5.** Theory of Agile Governance: conceptual framework. **Source**: Own elaboration.

As a result we have identified two different classes of system states during the theory building process: (1) **Macro-system states**: the system states related to the stage of awareness in agile governance; and, (2) **Micro-system states** (or plainly *system states*): the system states related to the operation of the theory.

---

[11] The criterion of inclusiveness refers to the need for all the units of the system to be included in the system state of the theory (Dubin, 1978; Torraco, 2000).
[12] The criterion of persistence requires that the system state persist through a meaningful period of time (Dubin, 1978; Torraco, 2000).
[13] The criterion of distinctiveness requires that all units take on determinant, that is, measurable and distinctive, values for the system state (Dubin, 1978; Torraco, 2000).

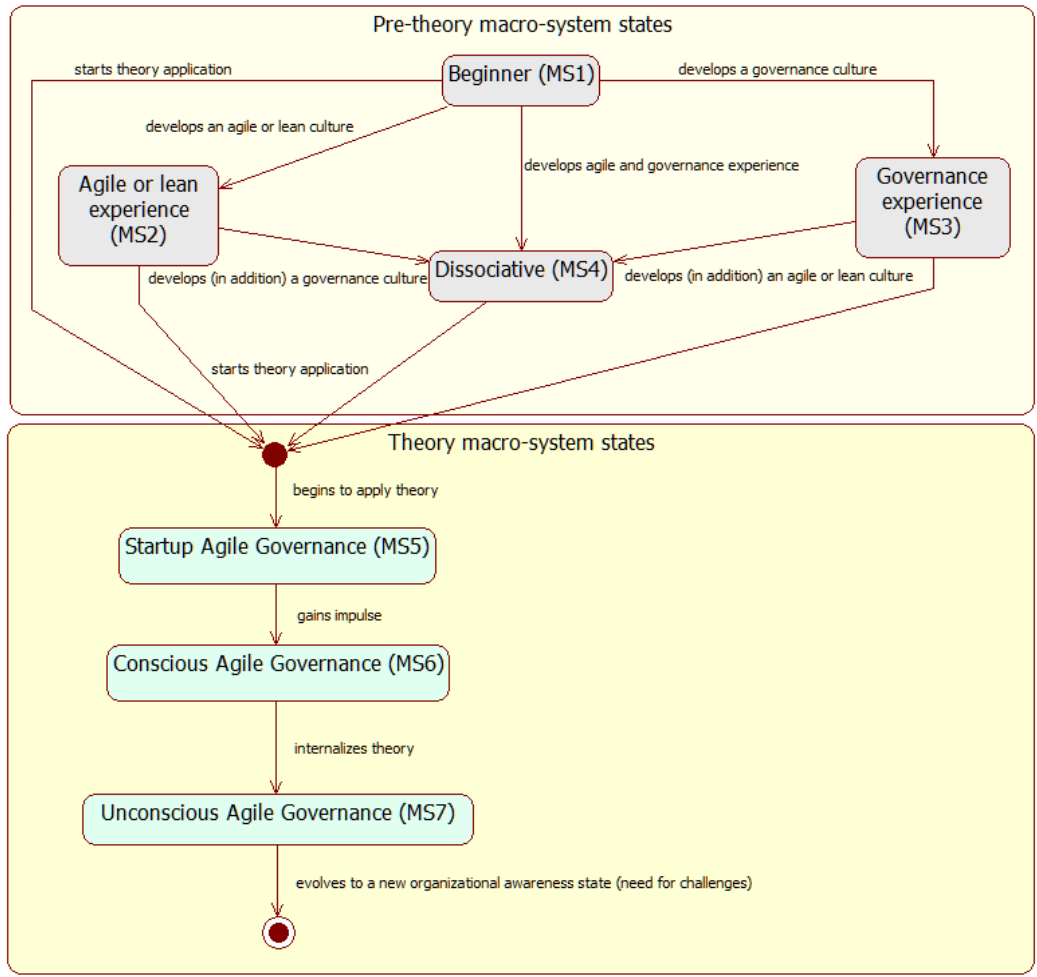

**Fig. 6.** Theory of Agile Governance: macro-system states. **Source**: Own elaboration.

Concerning to macro-system states, we have realized two types of them: (i) **Pre-theory macro-states**: related to the awareness found in the organizational context in the real world, before the theory application, whereas at least two of them were evidenced and discussed in (Luna et al., 2014), when we have highlighted overall trend movements in agile governance phenomena: **Trend 1** (agile or lean experience); and, **Trend 2** (governance experience); and, (ii) **Theory macro-system states**: related to the level of awareness in agile governance developed by means of the application of the theory. Those macro-system states are depicted in **Fig. 6**.

In short, the macro-system states are described as follows:

(MS1) **Beginner:** In this system state fits organizational contexts in which there is no governance experience, neither an agile culture established. This state is characterized by maximum values of [E] and [M], null values of [A] and [G], serendipitous values for [B], and minimum rate for [R] (likely very close to zero).

(MS2) **Agile or lean experience:** In this system state fits organizational contexts in which there is already an agile culture, however focused on specific agile approaches. They probably feel the need to implement governance practices. Occasionally, they wish to develop efforts to bring these practices to their core business. This state is characterized by high values of [E] and [M], null values for [G], and increasing values for [A] and [B] (likely low), as well as serendipitous values for [R] (likely low).

(MS3) **Governance experience:** In this system state fits organizational contexts in which there is already any governance experience. In some case, they perceive that the conventional practices can be heavy and/or bureaucratic. Once in a while, they wish to develop efforts to become governance quick and easy in order to achieve better results in their core business. This state is characterized by high values of [E] and [M], null values for [A], increasing values for [G] and [B] (likely low), as well as serendipitous values for [R] (likely low).

(MS4) **Dissociative:** In this system state fits organizational contexts in which there are already any specific agile approach and/or any governance experience (they may even have performed or be performing it), but they are not applying agile capabilities [A] and governance capabilities [G], in a combined and coordinated manner, to achieve better results in their core business. This state is characterized by high values of [E] and [M], as well as probably serendipitous values for [A], [G], [B], and [R] (likely low).

(MS5) **Startup Agile Governance:** In this system state fits organizational contexts in which has already started the application of the theory. This state is characterized by high (but decreasing) values of [E] and [M], as well as increasing values for [A], [G], [B] and [R] (likely low).

(MS6) **Conscious Agile Governance:** In this system state fits organizational contexts in which have already reached a primary level of organizational sustainability and competitiveness by application of the theory. This state is characterized by low (and decreasing) values of [E] and [M], as well as increasing values for [A], [G], [B] and [R] (likely high).

(MS7) **Unconscious Agile Governance:** In this system state fits organizational contexts in which have already reached a high level of organizational sustainability and competitiveness. They have already develop their activities in a high level of awareness (achieved by people and entire organizational context that have assimilated deeply the agile governance theory), acting and reacting in an unconsciously competent manner, almost intuitively, to deal with the emerging issues from the organizational context, as well as within the environment where they are inserted. This state is characterized by minimum values of [E] and [M] (likely very close to zero), maximum values for [A], [G], [B] and [R].

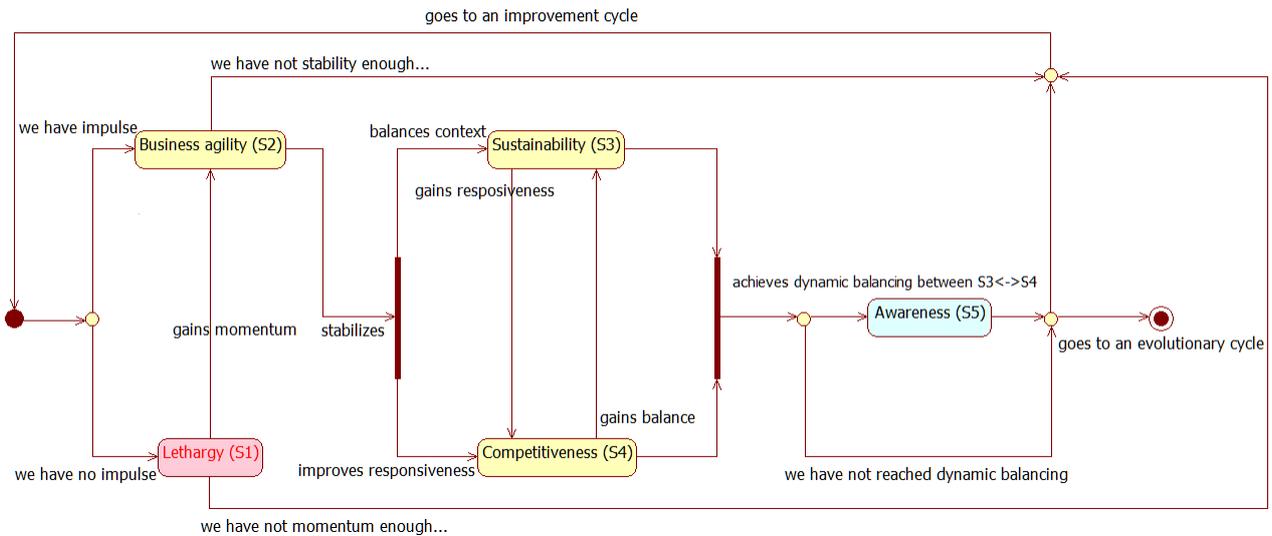

**Fig. 7.** Theory of Agile Governance: system states. **Source**: Own elaboration.

In complement, the system states related to the operation of the theory are depicted in **Fig. 7**. They occur within each macro-system state and are described as follows:

(S1) **Lethargy:** A lethargic state can compromise the entire organizational context, when fatigue, weariness (caused by exertion), or even by bad results or high level of stress caused by business pressure, befall upon the morale of the team. These circumstances entail on the following consequences: high (and increasing) values for [E] and [M] causing serendipitous (and likely very low) values for [A], [G], [B] and [R], which in turn generate their progressive decreasing.

(S2) **Business agility:** Business agility arises when the organizational context: (1) combines coordinately agile capabilities [A] and governance capabilities [G], applying subsequently their resultant effect upon business operations [B] (as described by 1$^{st}$ Law); or even, when, (2) agile capabilities [A] are applied directly on business operations [B] (as characterized by 2$^{nd}$ Law). The first approach entails the increasing of [A], [G] and [B], which in turn increases [R]; whereas the second approach keeps unchanged [G], but leads to the increasing of [A] and [B], which in turn enhances [R]. The effect of the former approach is *broader* and *systematic*, whereas the effect of the later approach is *localized* and *narrow*. Despite of the 1$^{st}$ Law generates faster results than 2$^{nd}$ Law, in both cases, respecting the proper proportions for each approach: [M] and [E] start to gradually decrease over the time, contributing to decrease the inhibition, restriction or disturbing on organizational context.

(S3) **Sustainability**: Organizational sustainability arises when [A] and [G] reach high values in the organizational context and their combined and coordinated application on [M], contributes to diminishing the inhibition and restriction [M] of the organizational context, even without changing significantly [E]. As a result, the gradual decreasing of [M] values accelerates the increase of [B], which in turn enhances [R].

(S4) **Competitiveness**: Organizational competitiveness emerges when [A] and [G] achieve high values in the organizational context and their combined and coordinated application on [E], contributes to decreasing the disturbances effects [E] felt by the organizational context, whereas causes a slight decreasing on [M]. As a consequence, the gradual reduction of [E] and [M] values speeds up the raising of [B], which in turn increases [R].

(S5) **Awareness:** Organizational awareness (or vitality) arises when the organizational context attains a responsive balancing by means of *sustainability* and *competitiveness* (i.e., a positive *dynamic balancing*[14] between these system states), resulting in a superior performance, where: (i) decreases to very low levels the influence of [E] and [M]; and, (ii) increases to very high grade the values for [A], [G], [B] and [R], which in turn cause their progressive and continuous increasing.

In alignment with Dubin's *inclusiveness* criterion, each of the units in the theoretical model is included and has a distinctive value in every system states. The emerged system states, also meets Dubin's additional criteria, namely: *determinate values* and *persistence*. In accordance with the determinate criterion, each of units within the theoretical model can be measured, at least in principle, during every system states. In accordance with the persistence criterion:

- The ***pre-theory macro-system states*** persist as long as the "time period" that the organizational context takes to adopt the theory or even, that an *unexpected event*[15] can take it to change to a different (pre-theory) macro-state.
- In turn, the ***theory macro-system states*** persist as long as the *agile governance evolutionary cycle* (as depicted in **Fig. 6**, and detailed in **Fig. 7**), taking into account how many *improvement cycles* that the organizational context need to achieve a new macro-system state.
- Finally, the ***micro-system states*** would persist as long as the agile governance *improvement cycle* occurs, as depicted in **Fig. 7**.

### 3.6  Organizational context and theory instantiation

When the *organizational boundary* (red dashed edge in **Fig. 5**) delimits the *internal environment*, separating it from the *external environment*, it characterizes the concept of **organizational context**. This concept works as a *control variable* of the theory. According to Creswell (2003), these variables are a special type of *independent variable* that are measured in a study because they potentially influence the *dependent variables*, i.e., a factor that strongly influences resulting values of the theory units, but it does not drive our theory. In other words, control variables could affect the values of the constructs, but it does not change the operating logic of the theory, neither the causality among the constructs (Creswell, 2003).

The **organizational context** can assume different values in our theory, such as: (1) teamwork; (2) project; (3) business unit; (4) entire enterprise; or even, (5) many institutions collaborating with each other in a multi organizational setting. We will refer those values as **levels of organizational context** according to their complexity: beginning the *teamwork* context as the lower level, and increasing gradually the complexity until reach the greater level of complexity, as the *multi organizational context*. In addition, the application of theory in each organizational context will be named **theory instance**.

For instance, the **Fig. 8** depicts an illustrative scenario, where as a matter of simplicity each theory instance was represented as an ***organelle***[16]. In other words, an *organelle* is a simplified version of the *conceptual framework of the theory* depicted in **Fig. 5**, as a streamlined schema of the theory, hiding the *constructs* and the *interactions* between them, but keeping the essential components to the discussions that follow.

---

[14] Referring to the adaptability of some "system states" with which a given "constructs' setting" may have a *stimulating effect* on the "organizational context" in one instance (*awareness* system state), and a soothing effect in another instance (*lethargy* system state).

[15] Any *unknown event* at the time of building of this theory, which the explanation or prediction is outside the scope of this theory.

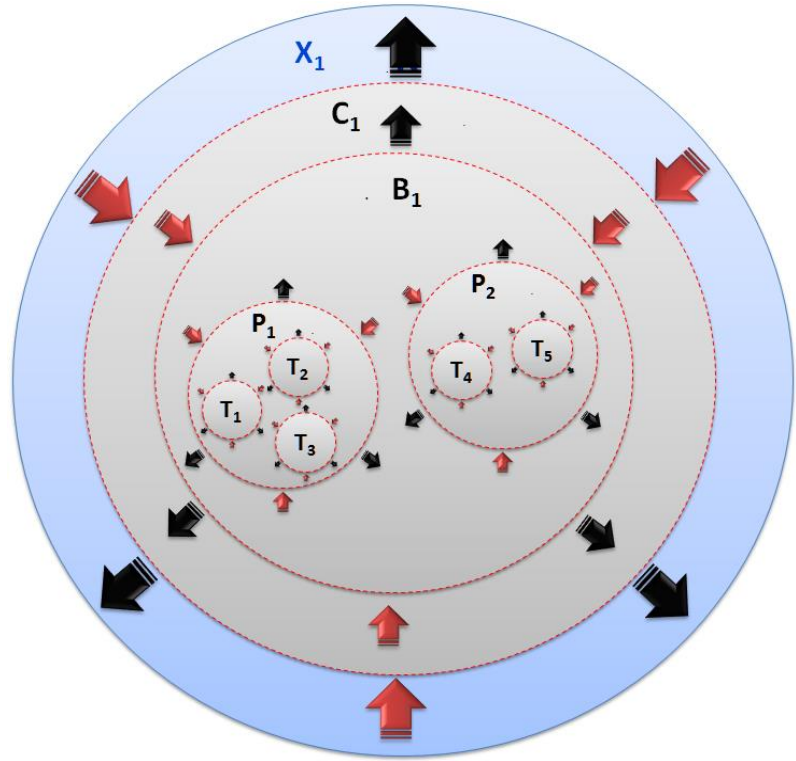

**Fig. 8.** Organizational context: multiple instances in a single enterprise. **Source**: Own elaboration.

On that scenario we can apply the theory in two different projects ($P_1$ and $P_2$) that belong to a same business unit ($B_1$), as well as apply the theory simultaneously to the business unit containing them ($B_1$). That business unit ($B_1$) is contained in a company ($C_1$), which in turn operates in a market ($X_1$).

In this case, the external environment ($E_{P_1}$ and $E_{P_2}$) to be considered for the theory application on two mentioned projects should be the environment of the business unit, ($E_{P_1} = (C_1 - P_1) \cup X_1$) and ($E_{P_2} = (C_1 - P_2) \cup X_1$), which containing them, while the external environment ($E_{B_1}$) to be considered for the business unit should be the company environment ($C_1$) where it is contained, i.e., ($E_{B_1} = (C_1 - B_1) \cup X_1$). We also, can consider that the project ($P_1$) is conducted by three teams ($T_1$, $T_2$ and $T_3$), while the other project ($P_2$) is carried out by other two teams ($T_4$ and $T_5$). Similarly, the identification of the external environment for each instance of the theory must be applied as done for $P_1$, $P_2$ and $B_1$.

It is inevitable to think that the most inner *organelles*, i.e., theory instance applied in a lower level of complexity, such as $T_1$, might be influenced by the disturbing factors from the external environment ($X_1$), as well as from the enterprise ($C_1$) in a diluted manner. Although other disturbing factors from the external environment of each level of organizational context which contains it ($P_1$, $B_1$ and $C_1$), can be added to the external disturbing resultant factors of the theory instance in question.

---

[16] A simplified manner to represent graphically a theory's instance.

For example, a sudden change in the exchange rate of a foreign currency, an external factor to the enterprise ($C_1$) from the market where it is inserted ($X_1$), can also affect a team ($T_1$). To make it happen, just that they have budgeted the cost of acquisition of some inputs (e.g., external software component or hardware device) for the project activities ($P_1$) in foreign currency, while they are billing the customer in local currency. Or even if they have subcontracted some service in foreign currency, although the project is being paid in local currency.

In each of these contexts the theory should be applied according the same general descriptions, but respecting the particularities of each organizational context. Moreover, we believe that the theory can be applied in a coordinated manner in different levels of organizational context, in a large number of possible combinations.

## 4    Conclusion

The outcome of the conceptual development phase of our theory-building research is a fully conceptualized theoretical model: *Theory of Agile Governance*. The components of the model are: the theory's *constructs*, its *laws of interaction*, its *boundary-determining conditions*, and its *system states*. Each of these components has been characterized and presented here.

We expect that the *conceptual framework* of the *Theory of Agile Governance* presented in this paper can provide some insights to understand the *agile governance phenomena* and consequently achieve the necessary fluency in this area of knowledge in order to bring it to a new level, accelerating its development, by scholars and practitioners.

As future work, we will carry out the second part of Dubin's method of theory building research: operationalize the conceptual framework of the theory and test its hypotheses by means of an empirical study. We are working to end up with a *trustworthy* theory to describe and analyze the agile governance phenomena, their constructs, mediators, moderators and disturbing factors.

**Acknowledgements**

We applied the SDC approach for the sequence of authors (Tscharntke, Hochberg, Rand, Resh, & Krauss, 2007). The authors acknowledge to CAPES, Brazil's Science without Borders Program, CNPq and ATI-PE by the research support. Special thanks to *Luciano José de Farias Silva,* and GPito team, namely: *Marcelo Luiz Monteiro Marinho*, *Robson Godoi de Albuquerque Maranhão*, *Suzana Cândido de Barros Sampaio,* and *Wylliams Barbosa Santos*; for their valuable contributions.